\documentclass[twocolumn,aps,pre,longbibliography]{revtex4-1}
\usepackage{amsmath,amssymb}
\usepackage{graphicx}% Include figure files
\usepackage{dcolumn}% Align table columns on decimal point
\usepackage{color}
\usepackage{epstopdf}
\usepackage[%
colorlinks=true,
urlcolor=blue,
linkcolor=blue,
citecolor=blue
]{hyperref}
\begin{document}
	\hyphenpenalty=5000
	%\tolerance=1000
	\hyphenation{ma-the-ma-ti-cal equi-li-brium sub-system tem-pera-ture eigen-levels distri-bution evo-lu-tion pro-ba-bi-li-ty tra-jectory macro-scopic micro-scopic quan-tum pro-perty ca-no-ni-cal sub-systems des-crip-tion tem-pe-ra-ture thermo-dynamics de-ge-ne-ra-cy thermo-dynamic phe-no-me-na phe-no-me-no-lo-gi-cal appli-cation theory Neumann eigen-energy react-ant ca-te-gor-iza-tion dif-ferent dif-ference patterns}
	\title{First principle thermodynamic study of oxygen vacancy at metal/oxide interface}
	\author{Guanchen Li}
	\email{guanchen@vt.edu}
	\author{Eric Tea}
	\author{Jianqiu Huang}
	\affiliation{%
		Department of Mechanical Engineering, Virginia Tech, Blacksburg, VA 24061, USA
	}%
	\author{Celine Hin}
	\affiliation{%
		Department of Mechanical Engineering, Virginia Tech, Blacksburg, VA 24061, USA\\
		Department of Materials Science and Engineering, Virginia Tech, Blacksburg, VA 24061, USA
	}%
	 
	\begin{abstract}
	The oxygen vacancy is a crucial intrinsic defect in metal-ultrathin oxide semiconductor heterostructures, and its formation at an interface is of great importance in determining the device performance and degradation. This paper presents an ab initio thermodynamic study of oxygen vacancies at metal/oxide interfaces. Electronic energies and entropies are calculated for defective interface systems, as a function of interface-vacancy distance. The study indicates that oxygen vacancies near the interface modify its bonding structure, and significantly change the thermodynamic properties of the system (i.e., electronic energy and entropy) compared to bulk-like oxygen vacancies. We illustrate that different factors, including the vacancy location dependence on the energy and entropy, the temperature dependence on the entropy, and the temperature and partial pressure dependence on the oxygen chemical potential, are all important in determining the Gibbs free energy of formation of oxygen vacancy.
	\end{abstract}
	\maketitle
\section{Introduction}
The design of many advanced functional materials relies on the precise control of point defects since they determine the material properties, and in some cases, the material failure \cite{Tuller2011,Bishop2011a,Bishop2011b}. Understanding and modeling such materials require a comprehensive knowledge of point defect characteristics, in particular, their thermodynamic properties (e.g., energy and entropy). Ab initio calculations based on Density Functional Theory (DFT) have become a very important tool to support the modeling of these defective materials \cite{Varvenne2013} and have succeeded in addressing various experimental questions relative to diffusion processes \cite{Ma2010,Wu2011}, phase transformations \cite{Hennig2005}, and the recovery of irradiated metals \cite{Fu2004,Fu2005,Schuler2015}.

The interface between materials is also of great interest, since heterostructures are widely used in energy conversion and electronic devices such as fuel cells \cite{Tuller2011}, solar cells \cite{Green2007}, capacitors \cite{Xie2012}, or metal-oxide-semiconductor devices \cite{Khan2000,Xie2012}. In these multi-layer structures, the interfaces between each components play a major role in determining various properties (e.g. electrical or thermal resistance), while the separate bulk contributions decrease. An interface breaks the translational symmetry of the bulk material in the direction perpendicular to the interface and results in a transition region that offers different stress conditions and different chemical environments to the atoms near the interface. Defect creation and accumulation at interface is likely to occur and can lead to device failure. For example, the dielectric breakdown, which is an old but still open problem, is believed to be related to the accumulation of defects at the metal-oxide interface \cite{Lombardo2005,Blochl1999,Usui2013}. Consequently, understanding how defects behave in presence of an interface, and their effect on the interface itself, is of great significance in predicting material performance.

Studies on defective interfaces are mainly conducted at the atomistic level, such as the calculations of Schottky barrier height, effective work function \cite{Tea2016,Kim2016},  and other material properties including formation and migration energy calculated at zero temperature \cite{Tang2007,Nadimi2010}. Indeed, temperature dependent properties (e.g., entropy, Gibbs free energy of formation) are rarely studied in a defective interface system at the atomic level using ab initio calculations. In this paper, we demonstrate the importance of both the temperature dependence, and the proximity of an interface, on oxygen vacancy properties in a metal/oxide interface system. The study carried on a typical metal/oxide interface, namely the Al/SiO$_2$ interface. At the atomic level, research on defect formation energy at an Al/SiO$_2$ interface is more complicated than at Si/SiO$_2$ interface \cite{Nadimi2010}, due to the presence of Al atoms. Oxygen atoms at the interface form two different types of bonds with Al and Si atoms. In a defective interface, both types of bonds can influence the thermodynamics of defects (i.e., vacancies or impurities) and determine the thermodynamic properties of the material (e.g., energy, entropy, and Gibbs free energy of formation). Taking advantage of ab initio calculations, comprehensive information on defective Al/SiO$_2$ interfaces will give an atomistic level picture on the oxygen vacancy behavior in the presence of metal/oxide interfaces.

The paper is organized as follows. In Section 2, we present the Density Functional Theory (DFT) calculation setup and then we introduce the thermodynamic framework for defect studies using Gibbs free energy. The Electron Localization Function (ELF) is used to show the bonds between the interface atoms. The band structure analysis offers information on the system density of states (DOS), enabling the identification of localized interface and defect states. In Section 3, we present the results on the formation energy (Section 3.1), electronic entropy (Section 3.2), and the correction to the formation energy from the electronic entropy (Section 3.3). In Section 4, we discuss the results. The paper is finished by the conclusion in Section 5.

\section{Methodology}
\subsection{System supercell and DFT calculations}
An Al/SiO$_2$ metal/oxide interface connecting Al (111) and SiO$_2$ (001) slabs is studied using supercell calculations (Fig. 1). The resulting lattice mismatch at the interface is less than $1$\% so that stress effects can be neglected. The supercell contains $7$ Al layers, 10 SiO$_2$ layers and a $\sim15$ {\AA} wide vacuum region separating periodic images. The interface plane contains 2x2 unit cells of SiO$_2$. The dangling bonds at the SiO$_2$ and Al open surfaces are passivated using hydrogen atoms and 0.75 fractional hydrogen atoms to lock the charges in their bulk-like configuration away from the metal-oxide interface. The defect-free interface system contains $224$ atoms in total (including $40$ Si, $80$ O, $84$ Al, $12$ H and $8$ 0.75H) and has been shown to be large enough to keep the interaction between the oxygen vacancy and its periodic image negligible \cite{Tea2016}. In order to catch the different thermodynamic features of the oxygen vacancy as a function of the distance from the interface, 10 different supercells are used. In each supercell, one oxygen atom is removed. The 10 supercells map 10 different oxygen vacancy locations (in the first $6$ oxygen layers) starting from the interface as shown in Fig. 1. For the same oxygen layer, the locations labelled with odd and even indices are not equivalent, due to the fact that the interfacial oxygen linked to Al are always on the right side of a silicon atom. In order to make a comparison between the density of states (DOS) calculated for all defective supercells, one helium atom is placed in the middle of the vacuum region to provide a common energy eigenlevel reference.

\begin{figure}[h]
	\centering
	\includegraphics[width=0.22\textwidth]{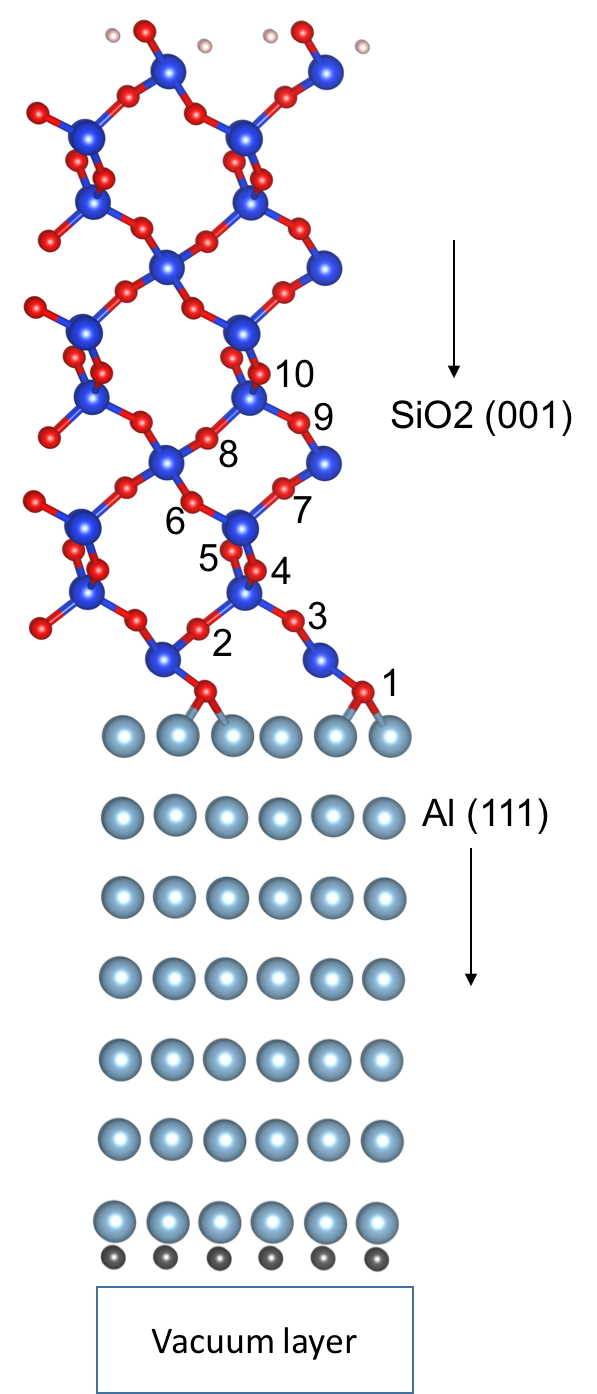}
	\includegraphics[width=0.22\textwidth]{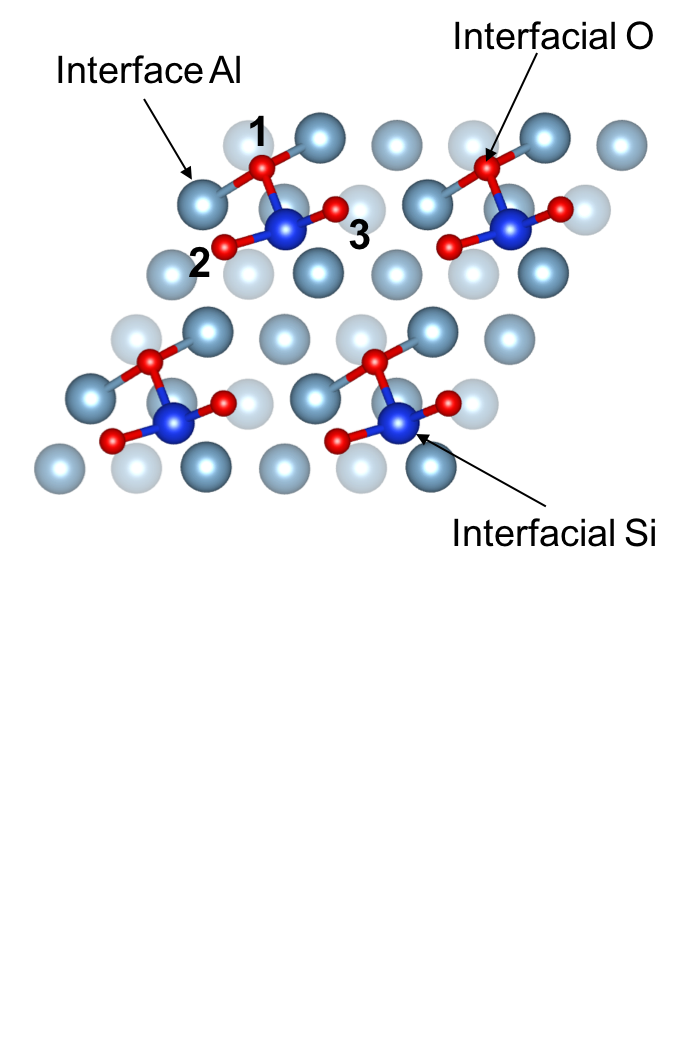}
	\caption{(Left) Sketch of the $\sim10$ {\AA} x$10$ {\AA} x$10$ {\AA} defect-free supercell, labelling the 10 oxygen vacancy locations before relaxation. The supercell contains $224$ atoms and a $\sim15$ {\AA} thick vacuum region. Surface passivating atoms, H and $0.75$ fractional H, are shown by pink and dark gray balls, respectively. In the density of states study, one Helium atom is placed in the middle of the vacuum region. (Right) Top view of the Al/SiO$_2$ interface showing the atomic layers around the interface ($1$ Si layer, $2$ oxygen layers and $3$ Al layers).}
	\label{fgr:structure}
\end{figure}
Ab initio calculations are performed in the framework of DFT. The Projector Augmented Wave method \cite{Blochl1994} is used as implemented in the Vienna Ab-initio Simulation Package \cite{Kresse1996a,Kresse1996b}. The Local Density Approximation (LDA) has been used for the exchange and correlation functional \cite{Perdew1981}. The plane wave cutoff energy is set to $400$ eV, and total energy convergence criteria to $10^{-7}$ eV. A 3x3x1 k-point mesh has been used. The Quasi-Newton algorithm has been employed to relax the atomic positions until forces on atoms are smaller than 5 meV$\cdot${\AA}$^{-1}$. This computational setup provides good convergence of the structures and total energies. Atomic structures and charge densities are visualized with VESTA \cite{Momma2008}.

\subsection{Defect thermodynamics}
\subsubsection{Gibbs free energy}
To model the thermodynamics of defective interfaces, especially their temperature dependent behavior, it is important to calculate the Gibbs free energy of formation, which is derived from the formation energy. The formation energy of a defect $X$ in charge state $q$ takes the form \cite{Walle1993,Zhang1991}:

\begin{equation}
E^f[X^q]=E[X^q]-E[perfect]-\sum_in_i\mu_i^0-qE_f
\end{equation}
where the first two terms on the right hand side are the total energies of a defective and a defect-free system. The positive (or negative) integer $n_i$ presents the number of atoms added (or removed) to form the defect, and $\mu_i^0$ is the corresponding chemical potential at zero temperature. In the calculation of formation energies, $\mu_i^0$ usually uses the energy of an isolated molecule at zero temperature. $E_f$ is the Fermi energy of electrons. The formation energy only contains the zero-temperature material properties, i.e., $E[X^q]$ and $E[perfect]$, which are calculated using DFT.

To take the non-zero temperature material properties into account, the Gibbs free energy of formation is used and is expressed as:
\begin{eqnarray}
G^f[X^q]=F[X^q]-F[perfect]-\sum_in_i\mu_i-qE_F
\end{eqnarray}
where $F=E-TS(V,T)$ is the free energy, and $\mu_i$ is the chemical potential. In our calculations, the volumes of defective and defect-free systems are assumed to be the same. Particularly, for the system with a neutral oxygen vacancy, the Gibbs free energy of formation takes the following form:
\begin{eqnarray}\label{eq:G}
G^f[V_O](P_{O_2},T)=F[V_O](T)-F[perfect](T)+\mu_O(P_{O_2},T)
\end{eqnarray}
where $\mu_O$ is the chemical potential of an oxygen atom, and $P_{O_2}$ is the oxygen partial pressure. The Gibbs free energy of formation is deduced from the free energies of defective and defect-free systems.

Electron and phonon both contribute to the free energy of a system. In this paper, we only focus on the electronic part of the free energy, which can be written as the sum of the electronic free energy at zero temperature $F^{el}_0$ and a temperature dependent term \cite{Freysoldt2014,Mermin1965}:
\begin{equation}\label{eq:F}
F(T)=F^{el}_0+\tilde{F}^{el}(T)
\end{equation}
$F^{el}_0$ is the total energy of the system at zero temperature. The temperature dependent term $\tilde{F}^{el}(T)$ is given by \cite{Kresse1996a,Methfessel1989}:
\begin{equation}\label{eq:f-t}
\tilde{F}^{el}(T)=-\frac{1}{2}TS^{el}+O(T^3)
\end{equation} 
Using Eq. (5), the temperature dependence can be taken into account by using the electron entropy $S^{el}$ calculated according to \cite{Eriksson1992,Wolverton1995}:
\begin{eqnarray}
S^{el}(T)=-k_B\int_{-\infty}^{+\infty}n(\epsilon,T=0)\{f(\epsilon,T)\ln f(\epsilon,T)\nonumber\\
+[1-f(\epsilon,T)]\ln[1-f(\epsilon,T)]\}d\epsilon
\end{eqnarray}
where $\epsilon$ is the energy, $f(\epsilon,T)$ is the Fermi distribution at temperature $T$, and $n(\epsilon,T=0)$ is the density of states at $T=0$.

The chemical potential of oxygen atoms $\mu_O$ brings in a temperature and a partial pressure dependence to the Gibbs free energy of formation. Unlike $F(T)$, $\mu_O$ is related to the chemical composition of defect, but is not material dependent. We calculate $\mu_O$ using \cite{Freysoldt2014}:
\begin{equation}\label{eq:chem}
2\mu_O(P_{O_2},T)=E_{O_2}+k_BT\left(\ln\frac{P_{O_2}V_Q}{k_BT}+\ln\frac{\sigma B_0}{k_BT}\right)
\end{equation}
where $E_{O_2}$ is the energy of an isolated oxygen molecule at zero temperature, $P_{O_2}$ is the partial pressure for O$_2$, $V_Q=(2\pi\hbar^2/mk_BT)^{3/2}$ is the quantum volume, $B_0$ is the rotational constant, and $\sigma$ is the associated symmetry factor ($2$ in the case of homonuclear diatomic molecules).

\subsubsection{Configurational entropy and defect areal concentration: interface supercell}
\begin{figure}
	\centering
	\includegraphics[width=0.48\textwidth]{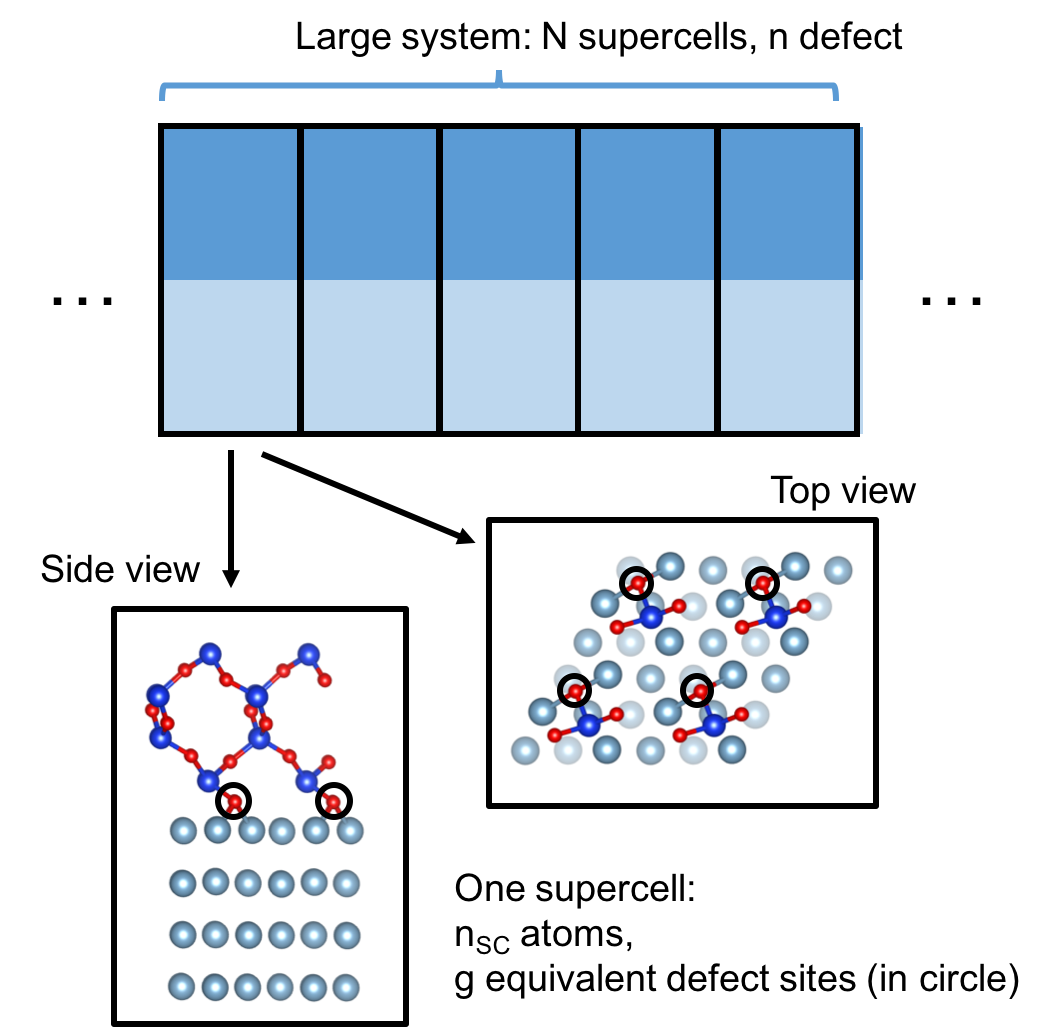}
	\caption{(Top) Sketch of a large system composed by $N$ supercells and n defects. (Bottom left) Side view of a supercell. (Bottom right) Top view of a supercell. In each supercell, the equivalent accessible sites for oxygen vacancies at location 1, i.e. in the first oxygen layer, are indicated by circles. The four sites per supercell in the first layer are equivalent, so $g=4$ for a defect at location 1.}
	\label{fgr:configurational entropy}
\end{figure}
The equilibrium concentration of defects can be calculated by adding the configurational entropy \cite{Keer1993} to the Gibbs free energy of formation. In particular, the configurational entropy here is derived from the atomic microstructure of the interface. As shown in Figure 2, one supercell contains $n_{SC}$ atoms and g accessible defect sites. A large system composed of $N$ supercells have therefore $gN$ accessible defect sites. If the total number of defects in the large system is $n$, the number of microstates $W$ to arrange those defects is:
\begin{equation}
W=C_{n}^{gN}=\frac{(gN)!}{(gN-n)!n!}\cong\frac{(gN)^n}{n!}
\end{equation}
Then, the defect configuration entropy in the large system is given by:
\begin{equation}
S^{conf}(n,N)=-k_B\ln W=k_B[n-n\ln(n/N)+n\ln(g)]
\end{equation}
and the corresponding total Gibbs free energy of the large system is:
\begin{eqnarray}
G(n,N)=NG[perfect]+nG^f-k_BT[c-c\ln c+c\ln g]\nonumber\\
\end{eqnarray}
where $c=n/N$ is the average number of defects per supercell, $G[perfect]$ is the Gibbs free energy of the defect-free supercell. By minimizing the average Gibbs free energy, the equilibrium value $c^{eq}$ equals to:
\begin{eqnarray}
c^{eq}=g \exp[-G^f/k_BT]
\end{eqnarray}
The areal concentration of defects is now given by $c^{eq}/A$ [unit: particle$\cdot$m$^{-2}$], where $A$ is the interface area of one supercell. Using Eq. (3)-(5), we can transform Eq. (11) to:
\begin{eqnarray}
-T\ln c^{eq}\propto G^f=(F_0^{el}-F_0^{el}[perfect])\nonumber\\
-S^{el}[perfect](T))+\mu_O(P_{O_2},T)
\end{eqnarray}
We define the zero-temperature vacancy formation energy using Eq. (1):
\begin{equation}
E^f = F_0^{el}[V_O]-F_0^{el}[perfect]+\mu_O^0
\end{equation}
where $\mu_O^0=0.5E_{O_2}$ is the oxygen atom chemical potential at zero temperature. To account for the temperature dependence, a correction to the free energy calculated from electronic entropy is defined by:
\begin{equation}
\Delta \tilde{F}^{el}(T)=-\frac{1}{2}T(S^{el}[V_O](T)-S^{el}[perfect](T))
\end{equation}
Thus, equation (12) takes the form:
\begin{equation}
-T\ln c^{eq}\propto G^f=E^f+\Delta \tilde{F}^{el}(T)+[\mu_O(P_{O_2},T)-\mu_O^0]
\end{equation}
In the result section, the three parts of the right hand side of Eq. (15) will be studied consecutively.

\subsection{Chemical bonds and band structure analysis}
To support our study on the vacancy formation energy, the electron localization function (ELF) \cite{Savin1997} has been used to provide detailed information about the chemical bonds at the interface. Moreover, the density of states (DOS) of defective interface systems for different interface-vacancy distances are compared to track the changes in interface and defect states. Special attention has been given to the bands near the Fermi level, which contribute the most to the electronic entropy. In addition to the DOS, the nature of the localized states (interface and/or defect) is of great interest as well, to elucidate the vacancy-interface distance dependence of the thermal properties. The projection of the wavefunctions on each atom evaluated at the $\Gamma$ point in reciprocal space is used as a basis for the determination of the localized nature of states. States localized on interface atoms are labeled as interface states, while those localized on the oxygen vacancy's neighboring Si atoms are labeled as defect states. As an exception, no defect state is labeled when the vacancy is at the first oxygen layer, since the interface and defect states are mixed together. By tracking the localized states, we acquire information on the coupling between the interface and defect states.

\section{Result}
In this section, calculation results for Al/SiO$_2$ supercells with different interface-vacancy distances are compared. To facilitate the discussion of the results, we introduce a notation; $V_O(x)$ stands for an oxygen vacancy at the location x varying from 1 to 10 as shown in Fig. 1. In Section 3.1, the oxygen vacancy formation energy, a zero-temperature material property, is studied. In Section 3.2, the electronic entropy, a non-zero temperature material property, is evaluated. In Section 3.3, we present the correction to formation energies brought by the inclusion of entropies.

\subsection{Electronic energy}
\begin{figure}
	\centering
	\includegraphics[width=0.48\textwidth]{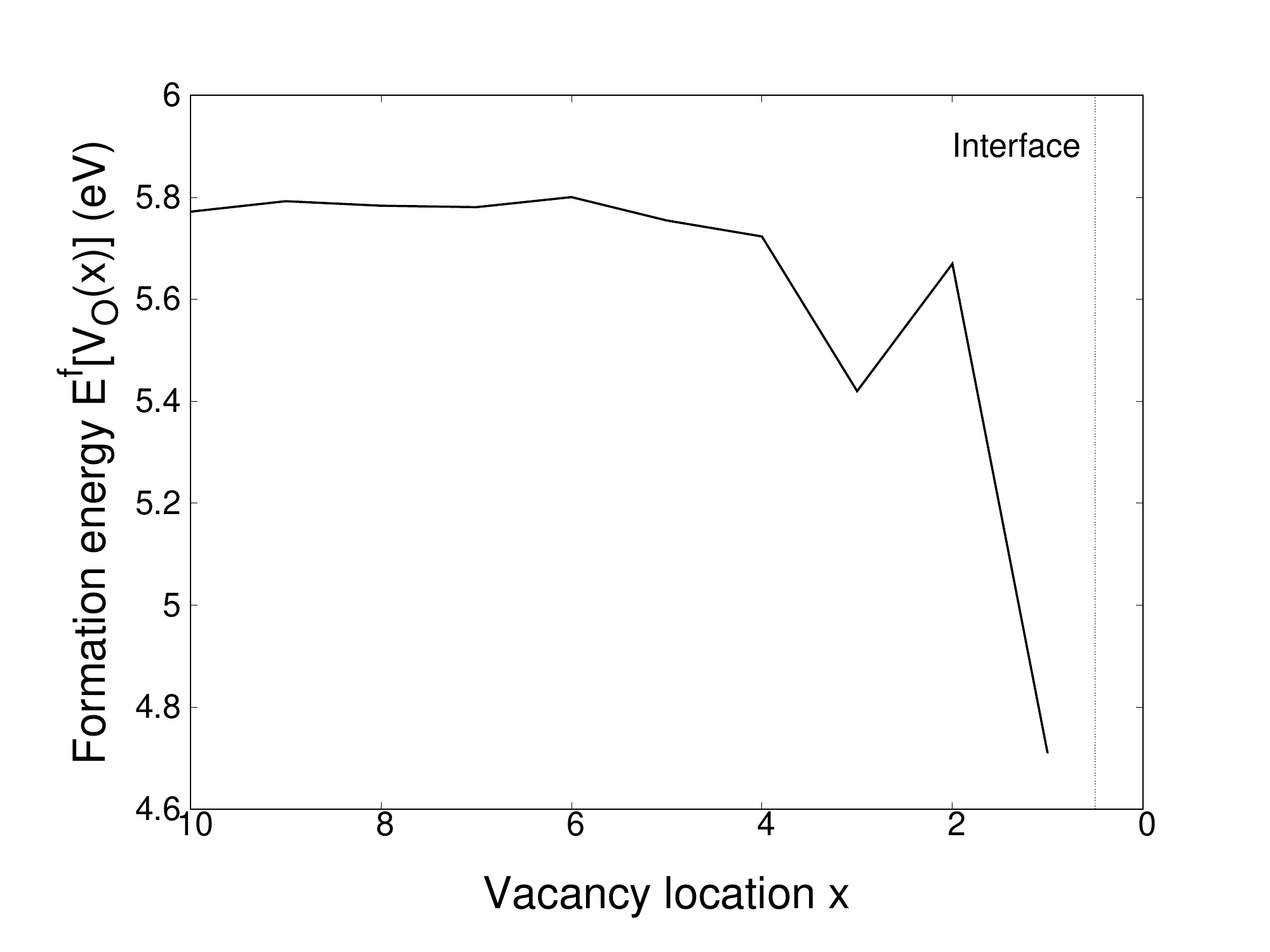}
	\caption{Formation energies of oxygen vacancies $E^f[V_O(x)]$ at different locations.}
	\label{fgr:Energy}
\end{figure}
In this section, we study the formation energies of oxygen vacancies $E^f[V_O(x)]$ as a function of their location. We use $\mu_O^0=0.5E_{O_2}$ as the oxygen chemical potential. The formation energies are shown in Fig. 3. In the bulk-like region far from the interface (sites 6 to 10), the vacancy formation energy is about $5.76\pm0.02$ eV {%, and the Si-Si bond length is $2.40${\AA}, 
in good agreement with the previous LDA calculations in bulk $\alpha$-quartz ($5.90$ eV% and $2.39${\AA}
\cite{Capron2000}, $5.4$ eV% and $2.393${\AA}
\cite{Roma2001}, and $5.80$ eV\cite{Martin2005}).} Ref. \cite{Sulimov2002} compiled formation energies varying between $6.5$ and $9.5$ eV, depending on the exchange and correlation functional and oxygen chemical potential used. However, the oxygen chemical potential is only an offset to the formation energy (see Eq. (1)). Since the present paper focuses on the relative formation energy, i.e. the difference between formation energies for vacancies at different locations, the chemical potential is not required.

In the first two layers of oxygen atoms from the interface (location 1, 2 and 3), the vacancy formation energies drop significantly. $E^f[V_O(3)]$, $E^f[V_O(2)]$, and $E^f[V_O(1)]$  are $0.35$ eV, $0.1$ eV, and $1.06$ eV lower than the bulk-like value $E^f[V_O(10)]$. While $V_O(2)$ and $V_O(3)$ belong to the same oxygen layer and have similar distances to the interface (see Fig. 1), $E^f[V_O(2)]$ is considerably higher than $E^f[V_O(3)]$ (higher by $\sim0.25$ eV). However, $V_O(2)$ and $V_O(3)$ are not equivalent with regard to the Al/SiO$_2$ interface system. As shown in Figure 1, the Al-O bonds always sit on the right side of the Si atom and make the oxygen atom at location 3 a little farther from the Al/SiO$_2$ interface after relaxation of the atomic positions. In order to elucidate this behavior, we plot in Figure 4 the ELFs near the interface to identify the interfacial bonding structures.

\begin{figure}
	\centering
	\includegraphics[width=0.22\textwidth]{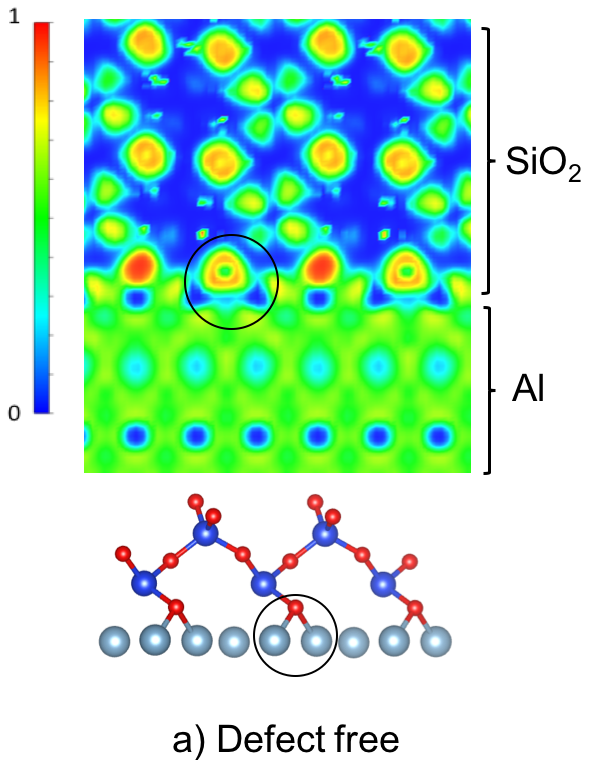}
	\includegraphics[width=0.22\textwidth]{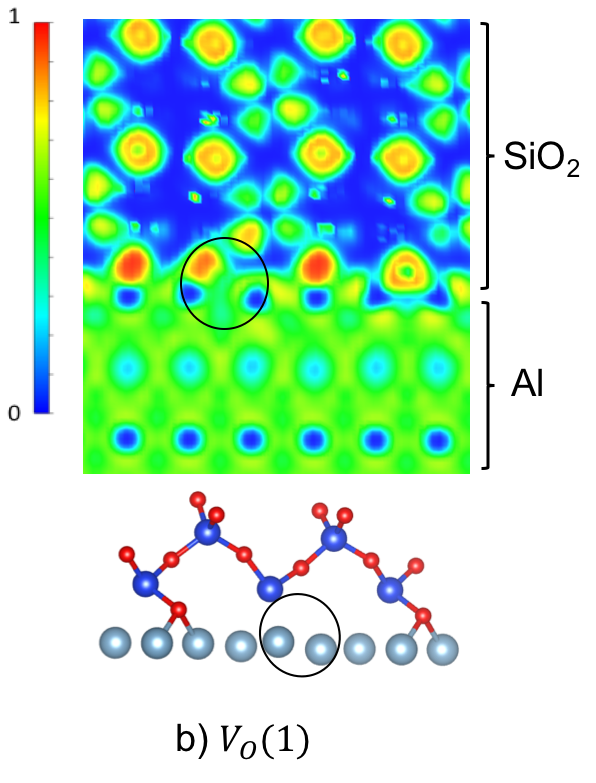}
	\includegraphics[width=0.22\textwidth]{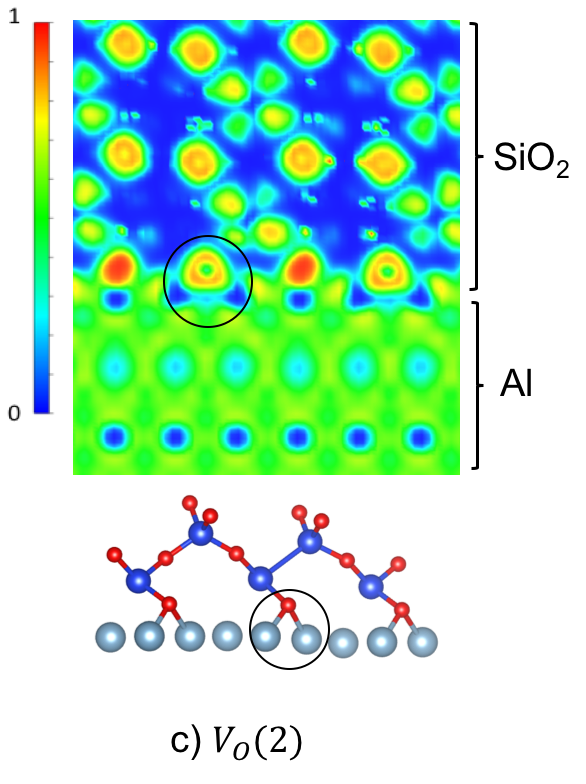}
	\includegraphics[width=0.22\textwidth]{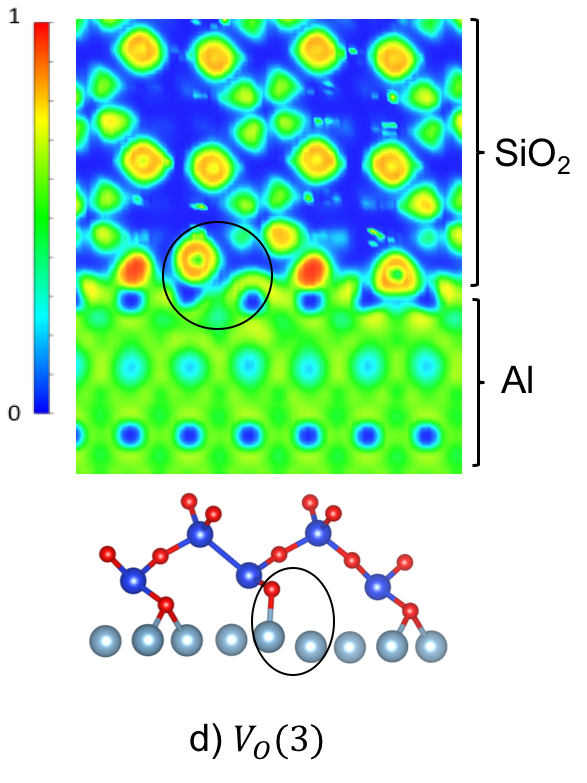}
	\caption{ELFs of the interfaces in the supercells of (1) defect free, (b) location 1, (c) location 2 and (d) location 3. The plane cut is decided by two Al atom and one O atom at the interface. The bottom part of the subfigure shows the corresponding atoms from the view of Fig. \ref{fgr:structure}.}
	\label{fgr:ELF}
\end{figure}

The ELFs provide information about charge localization and the bonding environment between atoms for different supercell configurations. In Fig. 4a, the ELF reveals the existence of an Al-O-Al structure at the defect-free interface and is used as a reference. For $V_O (1)$ (Fig. 4b), the Al-O-Al structure at the interface does not exist anymore, and the electrons are localized on the vacancy site (the red spot in the circle). For $V_O(2)$ (Fig. 4c), the Al-O-Al structure at the interface is preserved and retains an interface structure similar to the defect-free case one. For $V_O(3)$ (Fig. 4d), the vacancy is on the left side of the Si atom. We can observe that an Al-O bond is broken. The broken Al-O bond at the interface is responsible for the low vacancy formation energies for $V_O(1)$ and $V_O(3)$. The ELFs linked the bonding structure at the interface to the vacancy formation energy plotted on Figure 3.

\subsection{Electronic entropy: nonzero-temperature property}
\begin{figure}
	\centering
	\includegraphics[width=0.48\textwidth]{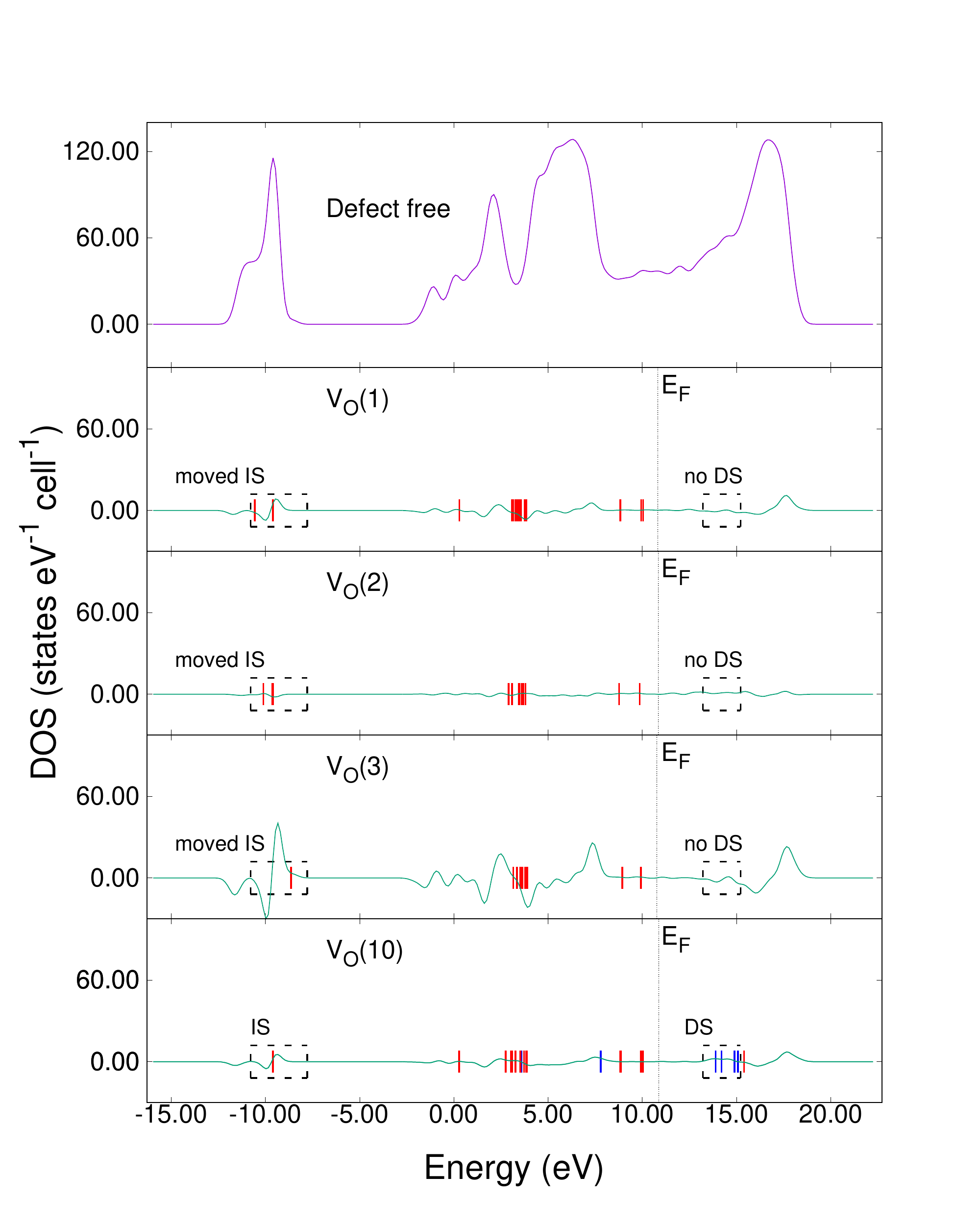}
	\caption{Comparison of density of states (DOS) of defective interfaces. The helium levels of the different systems are aligned at $0$ eV. The violet line in the top panel indicates the DOS of the defect-free interface. The green curves in the lower panels indicate the difference in DOS compared to the defect-free case. The vertical marks show the position of interface states (IS: red) and defect states (DS: blue). Dashed boxes indicate the energy regions of interest for the comparison of interface and defect states. }
	\label{fgr:DOS}
\end{figure}
The electronic entropy is a nonzero-temperature property of the system and another important contribution to the Gibbs free energy. According to Eq. (6), the DOSs of the defective supercells are used to derive the temperature dependence on the electronic entropy. In Fig. 5, the DOS of the defect-free case is used as a reference and is shown on the top panel. For $V_O(1)$, $V_O(2)$, $V_O(3)$ and $V_O(10)$, only the difference between the defect-free and defective DOS is plotted.

In the system with a bulk-like vacancy ($V_O(10)$), no significant change in the DOS compared to the defect free case is noticeable. Interface and defect states are clearly identified, which means that those two types of localized states do not mix with other states. In contrast, although $V_O(2)$ shows very little change in the DOS with respect to the defect free case as well, no defect states is observed and its interface states exhibit different energies. As an explanation, the interface configuration of $V_O(2)$ is very similar to the defect free interface, so the stress distribution in the material, the interfacial bonding structure, as well as the DOS, do not drastically change. However, due to the presence of vacancy in $V_O(2)$, the defect states are mixed with the interface states, so they cannot be clearly identified.

The DOSs of $V_O(1)$ and $V_O(3)$ exhibit considerable differences compared to the defect-free DOS. For these two cases, the broken Al-O bonds alters the atomic configuration at the interface and distort the stress distribution in the material. Particularly, $V_O(3)$ shows the largest difference in the DOS with respect to the defect-free case, which may substantially increase the entropy. Besides, less interface states compared to the bulk-like case are observed, and no defect state is found (right dashed black box in Fig. 5). Such a phenomenon implies that the wavefunctions localized at the interface mix with the localized states of the vacancy. More specifically, the localized interface states in the low energy region are relocated or split (left black boxes in Fig. 5). This can be explained by the atomistic reconfiguration of the interface as shown by the ELFs (Fig. 4).

\begin{figure}
	\centering
	\includegraphics[width=0.48\textwidth]{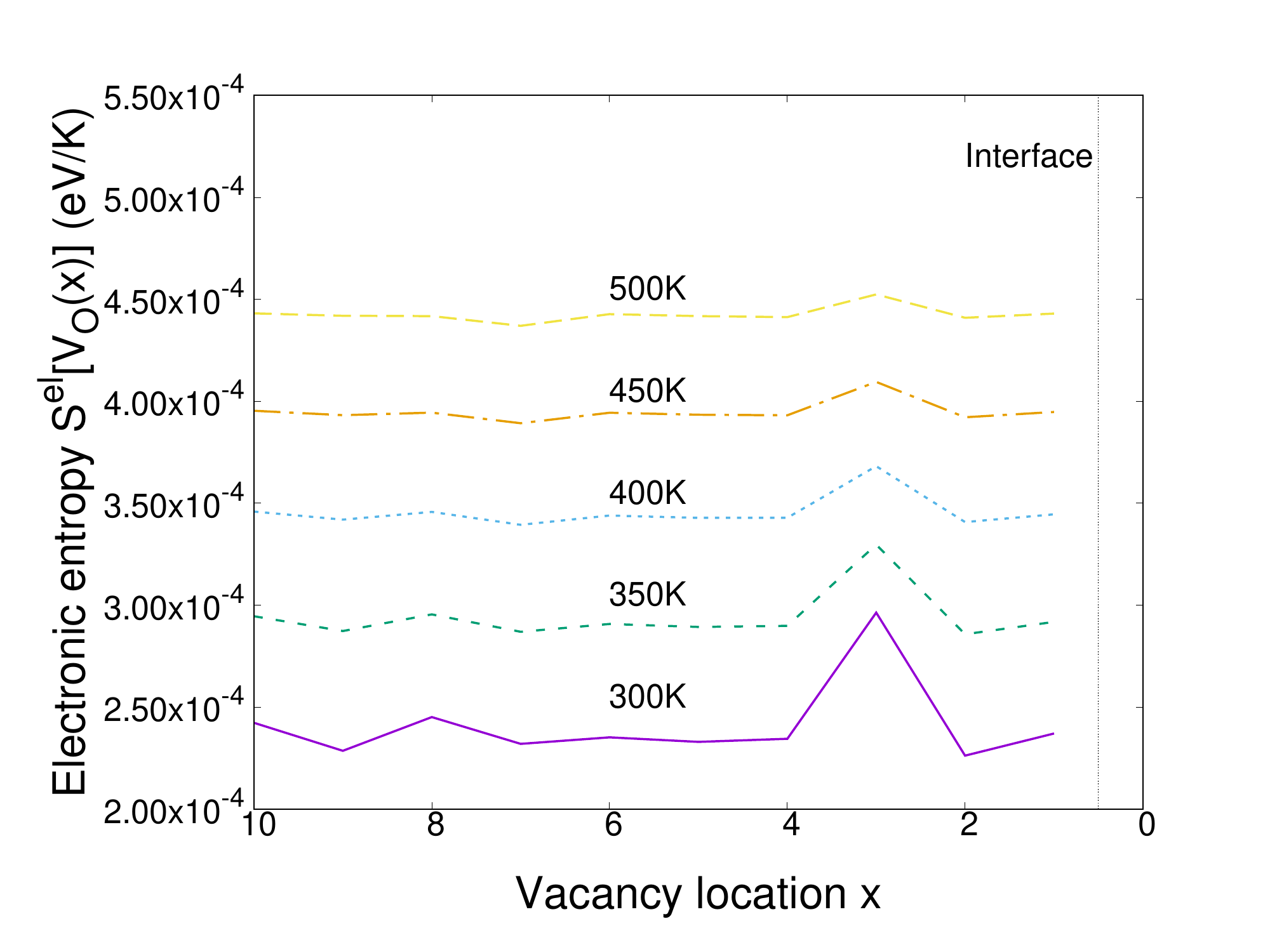}
	\caption{Electronic entropies of defective interfaces for different vacancy locations in the temperature range from $300$ K to $500$ K. }
	\label{S_T}
\end{figure}
In Fig. 6, the electronic entropies of defective supercells for different vacancy locations $S^{el}[V_O(x)]$ are plotted in the temperature range from $300$ K to $500$ K. At $500$ K, the vacancy location has nearly no influence on the electronic entropy. As the temperature decreases, according to the Fermi-Dirac electron distribution, less and less bands contribute to the entropy and a difference in entropy arises, especially when the vacancy is located at site 3. As expected from the DOS, $S^{el}[V_O(3)]$ shows significantly higher value compared to other vacancy locations. At $300$ K, $S^{el}[V_O(3)]$ is $3$x$10^{-4}$ eV/K, which almost equals the average entropy value at $350$ K. This result strongly impacts the Gibbs free energy of formation (see Section 3.3) and would consequently increases the vacancy concentration at location 3.

\begin{figure}
	\centering
	\includegraphics[width=0.48\textwidth]{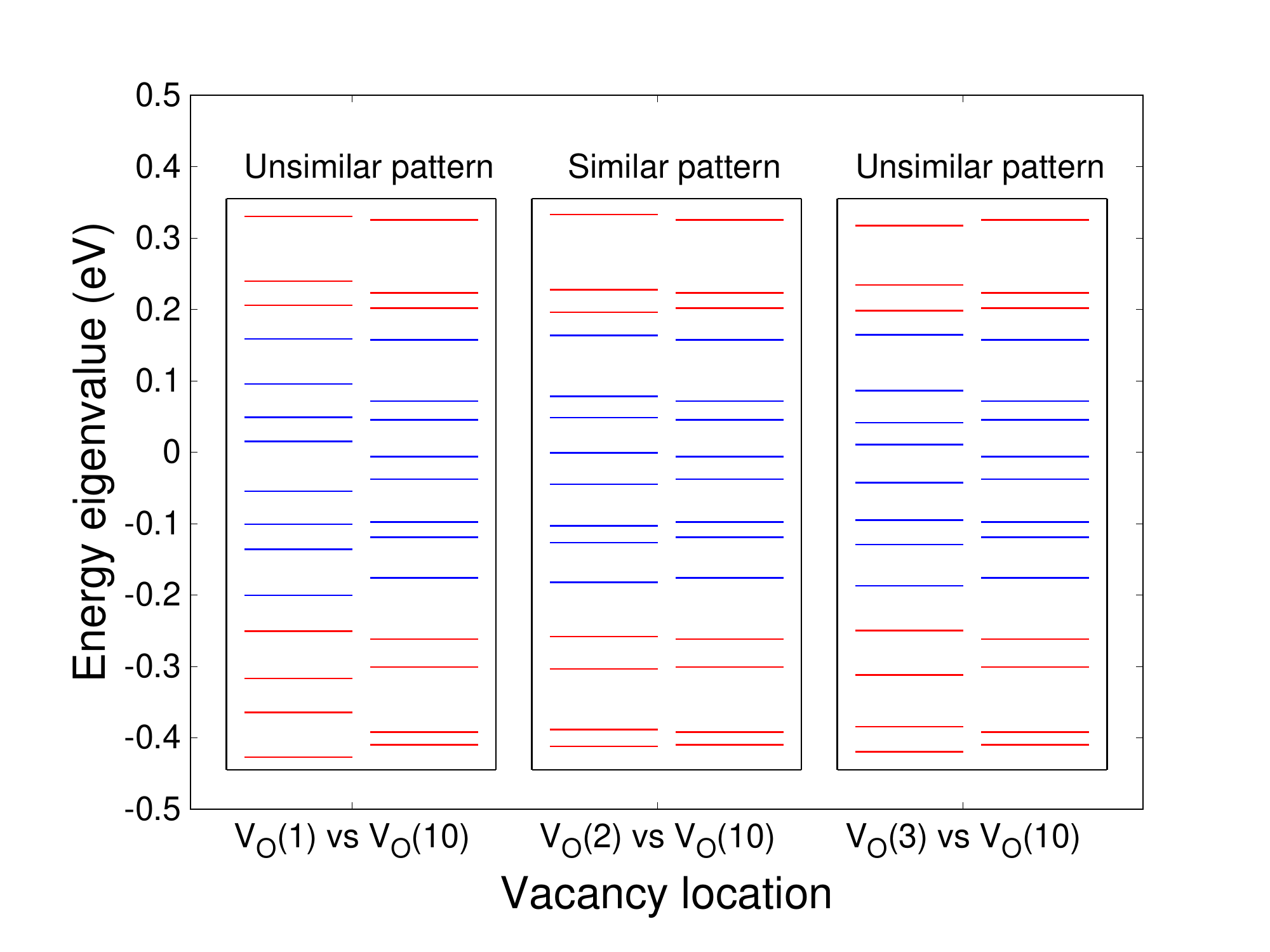}
	\caption{Energy levels of defective interfaces near the Fermi level (set to $0$ eV). The levels that contribute the most to the entropy at $300$ K and $500$ K are indicated by blue and red respectively.}
	\label{fgr:eigen}
\end{figure}
To provide a deeper understanding of the electronic entropy behavior, a close inspection of the energy levels near the Fermi levels is performed in Fig. 7. The Fermi levels of different systems are aligned at $0$ eV. The bands contributing the most to the entropy at $300$ K are drawn using blue lines, and red lines for $500$ K. The bands far away from the Fermi level are not shown, because their occupation probability takes either $1$ or $0$; so their contribution to the electronic entropy is negligible. The energy levels for $V_O(2)$ and $V_O(10)$ are comparable due to the similar Al-O-Al interface structures observed in both cases (see Fig. 4a and 4c). In contrast, the band structures for $V_O(1)$ and $V_O(3)$ exhibit many differences compared to $V_O(10)$, due to the changes in bonding at interface (see Fig. 4b and 4d). As the temperature increases from $300$ K to $500$ K, a broader range of eigenlevels around the Fermi level contribute significantly to the entropy. The differences (with respect to $V_O(10)$) in the entropy contributions from the blue region are balanced by those from the red region, which smears the entropy differences between different configurations of the systems at high temperature.

To summarize, the results extracted from the DOSs and energy levels near the Fermi levels clarify the influence of the band structures on the electronic entropies of the defective interface systems. More specifically, the DOS for $V_O(3)$ shows the largest difference compared to the defect-free interface case. The energy levels for $V_O(3)$ near the interface are also very different compared to those of the system with bulk-like vacancies ($V_O(10)$). As a result, $S^{el}[V_O(3)]$ has the largest value (see Fig. 6).

\subsection{Correction to formation energy using electronic entropies}
\begin{figure}
	\centering
	\includegraphics[width=0.48\textwidth]{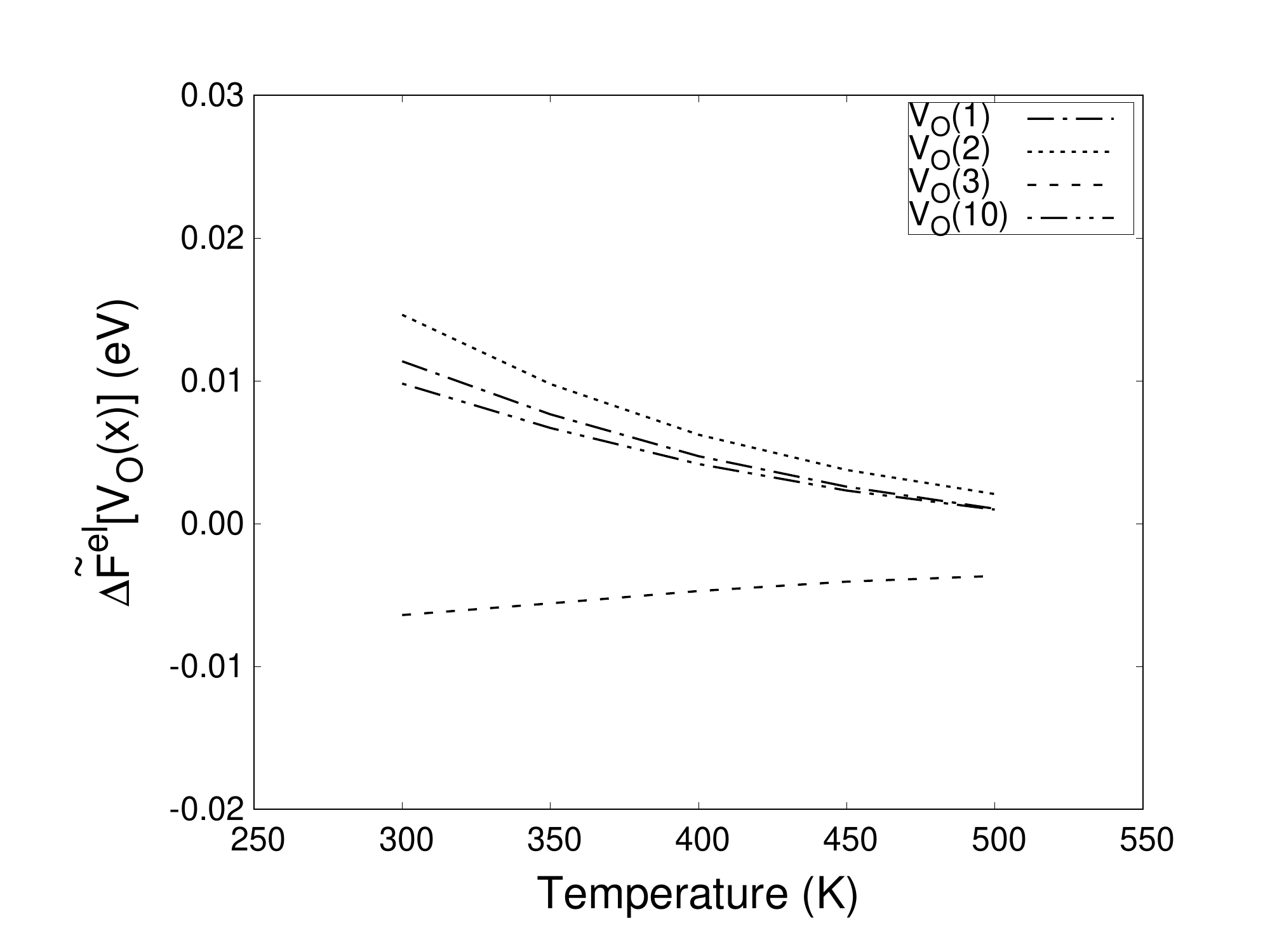}
	\caption{Corrections $\Delta \tilde{F}^{el}$ to the zero-temperature free energy, as a function of temperature and oxygen vacancy location.}
	\label{fgr:Correction}
\end{figure}
To get a proper estimation of the vacancy concentration for non-zero temperature, a correction $\Delta\tilde{F}^{el}$ (defined by Eq. (14)) to the zero-temperature formation energy is necessary to account for the temperature dependent behavior of the material energy and entropy. In Fig. 8, the correction $\Delta \tilde{F}^{el}$ is shown. The correction $\Delta \tilde{F}^{el}[V_O(3)]$ is negative, while it is positive for other locations. As a result, the correction makes $V_O(3)$ thermodynamically more favorable. At $300$ K, $\Delta \tilde{F}^{el}[V_O(3)]$ is about $0.02$ eV lower than $\Delta \tilde{F}^{el}[V_O(10)]$. This difference can double the vacancy concentration ratio between $V_O(3)$ and $V_O(10)$. As the temperature increases, the absolute values of $\Delta \tilde{F}^{el}$ decrease and become negligible ($\sim0.001$ eV) above $500$ K for all vacancy locations.

In Fig. 9, we highlight the environmental parameters that impact the most the vacancy concentration. More precisely, we aim at comparing the temperature and partial pressure dependence of $\mu_O$, along with the temperature dependence on $\Delta \tilde{F}^{el}$. We present the case where the vacancy is located on site 2, since Fig. 8 shows that $\Delta \tilde{F}^{el}[V_O(2)]$ has the largest absolute value and would change the concentration the most from the zero temperature value. As the temperature increases from 300 K to $500$ K, $\mu_O$ decreases by about $0.170$ eV, and the resulting vacancy concentration increases by a factor of $100$. For the same temperature range, the change in $\Delta \tilde{F}^{el}[V_O(2)]$ (dotted line) is only about $0.013$ eV, which would only double the zero temperature vacancy concentration. Consequently, the effect of temperature on $\mu_O$ will influence the vacancy concentration much more than its effect on $\Delta \tilde{F}^{el}[V_O(2)]$. However, the partial pressure dependence on $mu_O$ is comparable with the temperature dependence on $\Delta \tilde{F}^{el}[V_O(2)]$ in the $300$-$500$ K temperature and $10^4$-$10^5$ Pa partial pressure ranges. Indeed, as shown in Fig. 9, if  $P_{O_2}$ decreases by one order of magnitude at $300$ K, $\mu_O$ decreases by $0.03$ eV and would change the oxygen vacancy concentration by a factor of $3$.

\begin{figure}
	\centering
	\includegraphics[width=0.48\textwidth]{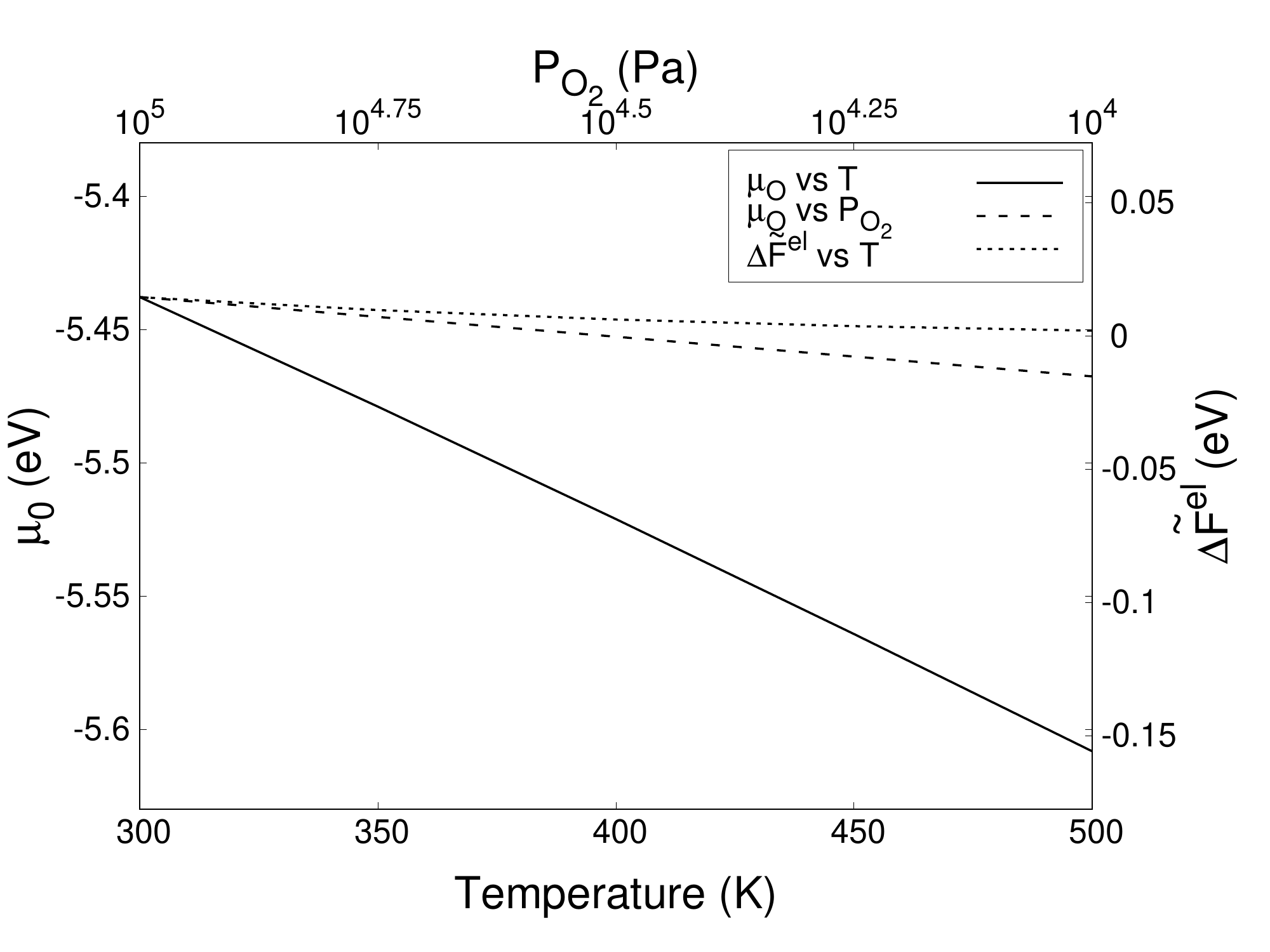}
	\caption{Comparison of the temperature dependence on $\Delta F^{el}[V_O(2)]$ (dotted line, x-axis: bottom, y-axis: right), the temperature dependence on $\mu_O$ (solid line, x-axis: bottom, y-axis: left), and the partial pressure dependence on $\mu_O$ (dashed line, x-axis: top, y-axis: left).}
	\label{fgr:Compare}
\end{figure}

In summary, the corrections $\Delta \tilde{F}^{el}$ calculated from electronic entropies are important at temperatures below $500$ K. For the vacancy concentration, the temperature dependence of $\Delta \tilde{F}^{el}$ is as important as the partial pressure dependence on $\mu_O$. These two effects have a smaller impact on the vacancy concentration than the temperature dependence on $\mu_O$. However, all of the three effects (i.e., the temperature dependence on the correction $\Delta \tilde{F}^{el}$, the temperature and partial pressure dependence on $\mu_O$) are not negligible and can influence the vacancy concentration by about $1$ to $2$ orders of magnitude.

\section{Discussion}
Based on the calculations presented in section 3, in this section we will discuss how this ab initio thermodynamic study on defective interface can advance the research of electronic devices, especially for thin layer devices. In section 4.1, we illustrate the influence of the vacancies near the interface on the dielectric breakdown and mechanical rupture of the device, using the results of vacancy formation energy as a function of vacancy locations (section 3.1). Then, in section 4.2, we clarify the importance of entropies in devices under operating conditions on device conception/design, using the results of section 3.2 and 3.3. 

\subsection{Vacancies near the interface}
In the first two oxygen layers, the vacancy formation energies drop significantly, compared to the bulk value. Consequently, oxygen vacancies are more likely to form in these two layers. Quantitatively speaking, the formation energies at locations 1 to 3 are $1.06$ eV, $0.1$ eV and $0.35$ eV lower than the bulk value $E^f[V_O(10)]$. At $300$ K, the corresponding equilibrium vacancy concentrations are about $6.4$x$10^{17}$, $48$ and $7.6$x$10^5$ times larger than the bulk value. Our calculations predict a non-uniform distribution of oxygen vacancies close to the interface. Ultra-thin oxides ($2.5\sim6$ nm), having a very narrow bulk like region, will be particularly impacted since 4 oxygen layers ($\sim0.5$ nm) can constitute a significant part of the total oxide thickness ($\sim20$\% for $2.5$ nm oxides). As the oxide thickness decreases in electronic devices, the leakage current will increase due to the tunnel effect. This effect can be amplified by the presence of interfacial oxygen vacancies which lower the tunneling barrier height \cite{Tea2016}, further increase the tunneling current, and ultimately triggering dielectric breakdown.

The percolation path model for dielectric breakdown qualitatively explained the lower density of defects required to trigger the breakdown in thin oxides, compared to thick oxides. The present study can advance the statistical approach in the percolation path model by accounting for spatially dependent vacancy formation energy instead of using an usual random defect generation scheme. The varying formation energy leads to a non-uniform distribution of the oxygen vacancies near the interface. Such a distribution will affect the prediction of conductive path formation compared to a uniform distribution.This more accurate picture of oxygen vacancy generation close to interfaces can provide more detailed information about dielectric breakdown especially in thin oxides where interfaces become prominent. 

Besides their impact on electrical properties, the oxygen vacancies in the first two oxygen layers also affect the system mechanical properties. The lower oxygen vacancy formation energies in the first two oxygen layers have been explained by the breakage of Al-O interfacial bonds. The broken Al-O bonds lead to a decrease of adhesion energy and distort the stress distribution near the interface. Both effects can create mechanical weak points at the interface, which could ultimately lead to rupture. 

\subsection{The importance of the entropy}
Our results suggest that more attention should be given to the entropy, which is usually neglected in first-principles studies for defect formation energies. The entropy, which is temperature dependent can strongly contribute to the point defect energetics. For a system operating under isothermal condition, a correction to the formation energy using the electronic entropy can give a better estimation of the vacancy concentration and subsequently a better prediction of material properties (e.g., electrical conductivity). At the Al/SiO$_2$ interface, the correction is considerable even at moderate temperatures ($<500$K). At $300$ K the correction is $\sim0.01$ eV, which affects the prediction of the vacancy concentration by $50$\%. The magnitude of the correction is comparable to the change in $\mu_O$ caused by reducing the oxygen partial pressure by 1 order of magnitude. As the temperature increases beyond $500$ K, the correction decreases and becomes negligible ($<0.001$ eV). 

In our calculations, only the electronic entropy has been accounted for. Besides the electronic entropy, the phonon entropy usually constitutes a very important contribution to the Gibbs free energy of the system. The configurational entropy of defects is also an important factor, although the configurational entropy of oxygen vacancies at different distance from the interface is the same in the present study. However, in other microstructures such as amorphous oxides, the different local environments around the vacancies can significantly change the configurational entropy and ultimately impact the Gibbs free energy of formation. This further stresses the importance of entropic contributions in point defect energetics.

Finally, it will be even more important in device conception and design to account for entropy. Under operating conditions, the devices are not working under isothermal conditions even in the presence of a cooling system. The local heat release during operation can result in hot spots and a non-uniform temperature distribution in the device. From the thermodynamic viewpoint, the prediction of vacancy concentrations under isothermal conditions is only valid when the system keeps in a mutual equilibrium with an external reservoir and has very slow temperature change rate. However, if the cooling speed is much slower than the heat release in a specific region, that region will undergo an adiabatic process rather than an isothermal one. In this case, the maximum entropy principle gives the equilibrium state, and the role of the Gibbs free energy in determining the vacancy concentration is less important. In the present study, the vacancies at the location 3 will become even more thermodynamically favorable under adiabatic conditions due to their higher entropy. Thus, when interface regions in a device cannot be cooled down fast enough, the heat release (due to Joule heating or reaction heat of the vacancy generation) can enhance the generation of vacancies close to the interface, which could ultimately lead to the dielectric breakdown or mechanical failure of the devices.

\section{Conclusion}
This paper provides a thermodynamic analysis of a defective interface system (neutral oxygen vacancy at Al/SiO$_2$ interface) using ab initio DFT calculations and a thermodynamic framework. We found that the vacancy location impacts the atomistic structure of the interface, and consequently the electronic and thermodynamic properties (i.e., electronic energy and entropy). More specifically, the vacancy creation near the interface can break interfacial Al-O bonds, change the DOS and electronic entropy, and finally lead to an increase in the Gibbs free energy of formation. Another important result is the temperature and partial pressure dependence of the Gibbs free energy of formation. Our study helped linking the existing zero-temperature ab initio study of defective interfaces to non-zero temperature experiments under certain temperature and partial pressure conditions. The vacancy concentration can vary by $1$-$2$ orders of magnitude within the $300$-$500$ K temperature range and $10^4$-$10^5$ Pa partial pressure range. These joint calculations can also enhance our understanding of phenomena happening at defective interface. Taking into account both the inhomogeneous vacancy concentration and the entropy in more advanced transport models will support the research on the failure of thin layer devices, especially when the heat release is significant.

\section{Acknowledgements}
This work was funded by the Air Force with program name: Aerospace Materials for Extreme Environment and grant number: FA9550-14-1-0157. We acknowledge Advanced Research Computing at Virginia Tech for providing the necessary computational resources and technical support that have contributed to this work (http://www.arc.vt.edu).

\bibliography{pccp}
\bibliographystyle{rsc} %the RSC's .bst file
\end{document}